\documentclass{jaa}
\usepackage[columnwise,switch]{lineno}
\usepackage{graphicx}
\usepackage{xspace}	

\usepackage{threeparttable}
\usepackage{amsmath}
\usepackage[utf8]{inputenc}
\usepackage{longtable}
\usepackage{graphicx}
\usepackage{natbib}
\usepackage{multirow}
\usepackage[Table,xcdraw]{xcolor}
\usepackage{rotating}
\newcommand{\apj}{ApJ}

\newcommand{\mnras}{MNRAS}

\newcommand{\physrep}{Physics Reports}
\newcommand{\nat}{Nature}
\newcommand{\apjl}{ApJL}
\newcommand{\apjs}{ApJS}

\newcommand{\aap}{Astronomy and Astrophysics}

\newcommand{\aj}{AJ}

\newcommand{\ssr}{Space Science Reviews}

\newcommand\T{\rule{0pt}{2ex}} 

\newcommand{\tninty}{{$T_{\rm 90}$}\xspace}
\newcommand{\thisgrb}{GRB~210217A\xspace}
\newcommand{\swift}{{\it Swift}\xspace}
\newcommand{\swiftT}{{T$_{\rm 0}$}}
\newcommand{\fermi}{{\it Fermi}\xspace}
\newcommand{\gbm}{{\it Fermi}-GBM }
\newcommand{\bat}{{\it Swift}-BAT }
\newcommand{\xrt}{{\it Swift}-XRT }
\newcommand{\uvot}{{\it Swift}-UVOT }

\newcommand{\keV}{{\rm keV}\xspace}

\newcommand{\mvts}{{$t_{\rm mvts}$}\xspace}
\newcommand{\lmin}{{$\Gamma_{\rm min}$}\xspace}
\newcommand{\Ep}{$E_{\rm p}$\xspace}
\newcommand{\sw}[1]{\texttt{#1}}
\usepackage{lipsum}

\usepackage{hyperref}
\hypersetup{colorlinks=true,
    breaklinks=true,
    urlcolor= black,
    linkcolor= blue,
    bookmarksopen=false,
    filecolor=black,
    citecolor=blue,
    linkbordercolor=blue
}

\begin{document}\sloppy
%\linenumbers
\title{\thisgrb: A short or a long GRB?}

\author{Dimple\textsuperscript{1,2,*}, Kuntal Misra\textsuperscript{1}, Ankur Ghosh\textsuperscript{1,3}, K. G. Arun\textsuperscript{4}, Rahul Gupta\textsuperscript{1,2}, Amit Kumar\textsuperscript{1,3}, L. Resmi\textsuperscript{5}, S. B. Pandey\textsuperscript{1} and Lallan Yadav\textsuperscript{2}}
\affilOne{\textsuperscript{1}Aryabhatta Research Institute of Observational Sciences (ARIES), Manora Peak, Nainital-263002, India.\\}
\affilTwo{\textsuperscript{2}Department of Physics, Deen Dayal Upadhyaya Gorakhpur University, Gorakhpur-273009, India.\\}
\affilThree{\textsuperscript{3}School of Studies in Physics and Astrophysics, Pt. Ravishankar Shukla University, Chattisgarh 492010, India.\\}
\affilFour{\textsuperscript{4}Chennai Mathematical Institute, Siruseri, 603103 Tamilnadu, India.\\}
\affilFive{\textsuperscript{5}Indian Institute of Space Science and Technology, Trivandrum 695 547, India.\\}

\twocolumn[{

\maketitle

\corres{dimple@aries.res.in}

\msinfo{2021}{---}

%%abstract
\begin{abstract}
Gamma-ray bursts are traditionally classified as short and long bursts based on their \tninty value (the time interval during which an instrument observes $5\%$ to $95\%$ of gamma-ray/hard X-ray fluence). However, \tninty is dependent on the detector sensitivity and the energy range in which the instrument operates. As a result, different instruments provide different values of \tninty for a burst. \thisgrb is detected with different duration by \swift and \fermi. It is classified as a long/soft GRB by \bat with a \tninty value of 3.76 sec.
On the other hand, the sub-threshold detection by \gbm classified \thisgrb as a short/hard burst with a duration of 1.024 sec. We present the multi-wavelength analysis of \thisgrb (lying in the overlapping regime of long and short GRBs) to identify its actual class using multi-wavelength data. We utilized the \tninty-hardness ratio, \tninty-\Ep, and \tninty-\mvts distributions of the GRBs to find the probability of \thisgrb being a short GRB.
Further, we estimated the photometric redshift of the burst by fitting the joint XRT/UVOT SED and place the burst in the Amati plane. We found that \thisgrb is an ambiguous burst showing properties of both short and long class of GRBs.
 
\end{abstract}

\keywords{gamma-ray burst: general, gamma-ray burst: individual: \thisgrb, methods: data analysis}
}]
%Bayesian Gaussian mixture model

\doinum{}
\artcitid{\#\#\#\#}
\volnum{000}
\year{0000}
\pgrange{1--}
\setcounter{page}{1}
\lp{11}

\section{Introduction}

Gamma-ray bursts (GRBs) are the short and intense pulses of $\gamma$-rays, occurring randomly at a rate of $\sim$ 1 event per day. The bi-modality in \tninty distribution of GRBs is used to divide these energetic events into two classes: short and long bursts with a boundary at 2 sec \citep{1993ApJ...413L.101K}. However, the \tninty value relies on the energy range in which the instrument operates and its trigger method. In general, bursts have a lower value of \tninty in higher energy channels \citep{2013MNRAS.430..163Q, 1995ApJ...448L.101F}. The \tninty value also depends on the sensitivity of the detector and background variations. Further, the observed \tninty depends on the redshift; the rest frame duration (\tninty/(1+$z$)) will be lower than the observed. It is challenging to decide the class of the GRB without any redshift information. It is also observed that some GRBs with \tninty values favourable to long GRBs have afterglow and host properties similar to the short GRBs \citep{2006Natur.444.1053G}. On the other hand, other GRBs with \tninty less than 2 sec show properties identical to long GRBs \citep{2009A&A...507L..45A, 2021NatAs.tmp..142A}. 

Therefore, it is difficult to classify the GRBs based on \tninty alone, particularly for GRBs lying close to the boundary. It is also essential to look for other observational characteristics apart from their \tninty information that can distinguish the two classes. 

Hardness ratio (HR), the fluence ratio in harder to softer energy bands,  can be used to classify GRBs. Short GRBs are found to be harder with large values of HR compared to their long companions \citep{1998ApJ...497L..21T, 1995ARA&A..33..415F}. Furthermore, HR is found to be correlated with \tninty for the complete sample of GRBs. However, no correlation is noticed between the two for an individual class \citep{2000PASJ...52..759Q}. 

Another characteristic is the spectral lag (i.e., the delay in the arrival times of low-energy photons to high-energy photons) which can differentiate the two classes. Long GRBs have significant lags (up to a few seconds for some of them) in their light curves in different energy channels. On the other hand, no lag (nearly zero) is observed for short GRBs \citep{1995A&A...300..746C, 2006MNRAS.367.1751Y, 2015MNRAS.452..824K}.

The two classes can also be compared concerning their energetics ($E_{\rm \gamma, iso}$) and luminosities ($L_{\rm p, iso}$). Short GRBs have, on average, energies that are smaller than that of long bursts \citep{2009A&A...496..585G}. They are located at two different places in $E_{\rm \gamma, iso}$ -\Ep plane (Amati plane), following a correlation that is parallel to each other \citep{Amati_2002, Amati_2006}. Additionally, the host galaxy properties such as stellar population, star formation rate (SFR), morphology, offset, etc., are different for short and long bursts in general and can provide a clue about the burst progenitor system and hence the class of the GRB \citep{Li_2016}.

\begin{figure}[!ht]
\includegraphics[scale=0.32]{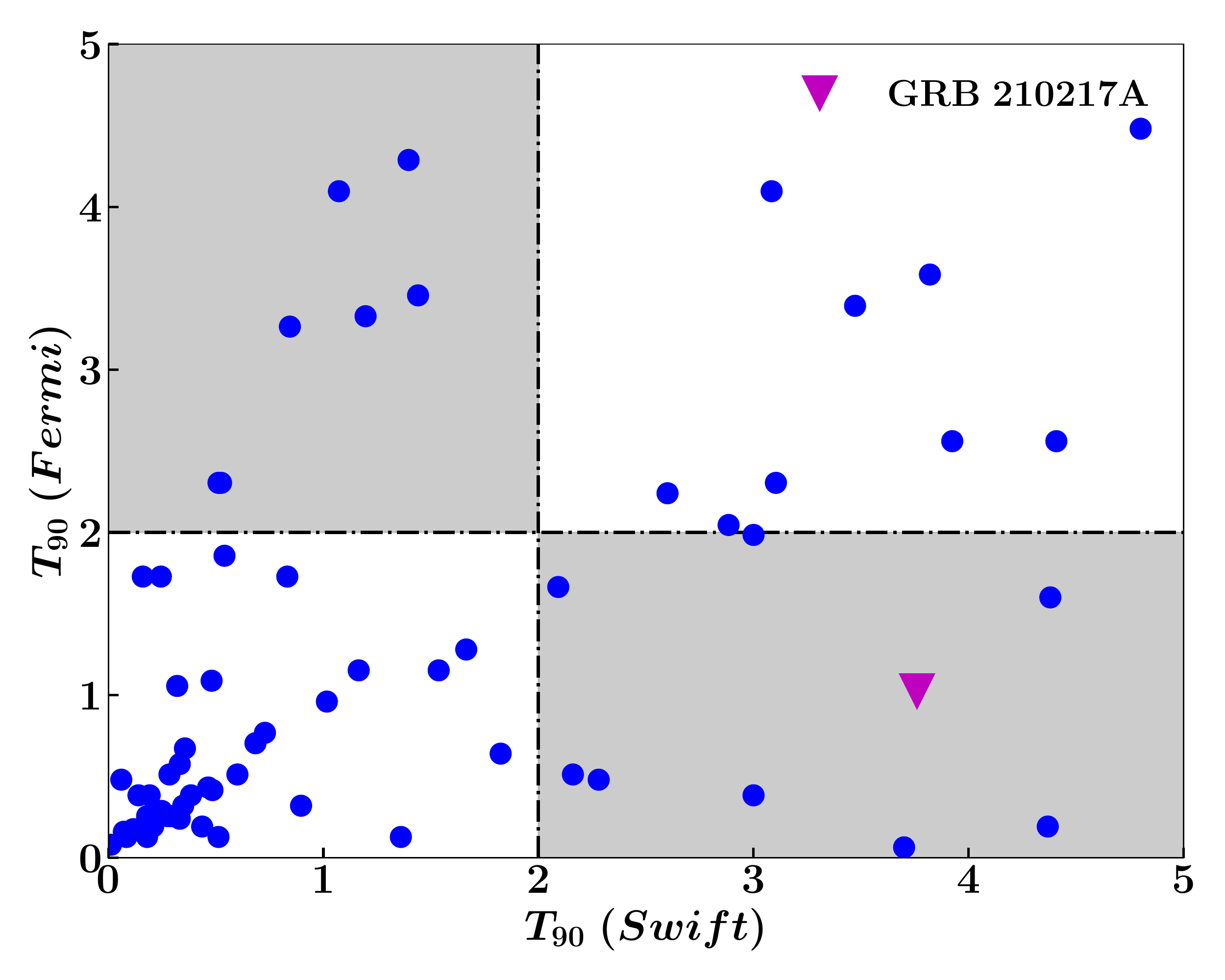}
\caption{\tninty duration of GRBs detected by both \fermi (50-300 \keV) and \swift (15-350 \keV). GRBs lying in the first and third quadrant have the same classification from both satellites. However, GRBs lying in the second and fourth quadrant (shaded grey region) have different classifications from both satellites. The magenta triangle represents \thisgrb. }
\label{sample}
\end{figure} 

The sensitivity of detectors affects the duration timescales of the bursts. As a result, GRBs detected by \swift and \fermi have different values of duration reported. Figure \ref{sample} represents the \tninty values as measured by \swift(15-350 \keV) and \fermi (50-300 \keV). The bursts have different values of \tninty reported by \swift and \fermi. Most of them are lying in the same class, either short or long burst (first and third quadrant). There are 16 GRBs detected by \swift and \fermi between 2008 to February 2021, including \thisgrb (second and fourth quadrant, the shaded grey regions), having different classification provided by two satellites.

GRB 210217A is one of the recent bursts lying at the boundary of short and long GRBs divide with different burst duration values reported by \swift and \fermi. \thisgrb was detected by \bat \citep{2021GCN.29521....1S} and \gbm \citep{2021GCN.29536....1F}. \swift Burst alert telescope (BAT) reported a \tninty of $4.22\pm1.15$ sec (15-350 \keV), suggesting it to be a long GRB \citep{2021GCN.29534....1S}. The sub-threshold detection by \gbm with a duration of 1.024 sec (25-294 \keV) suggests that the burst might belong to the short population of GRBs\footnote{\url{https://gcn.gsfc.nasa.gov/notices_gbm_sub/635297147.fermi}}. This burst was well within the observational capabilities of moderate-sized Indian telescopes. Therefore, we observed the optical afterglow of this burst with the 1.3m Devasthal Fast Optical Telescope (DFOT) and 3.6m Devasthal Optical Telescope (DOT) located in Aryabhatta Research Institute of Observational Sciences (ARIES), Nainital. The study of prompt and afterglow emission of a GRB provides a complete picture of the nature of the GRB.

We performed a detailed analysis to classify \thisgrb. This paper presents the multi-wavelength analysis of \thisgrb and discusses various methods for classification. Further, we estimated the burst's redshift using joint fitting of the spectral energy distribution (SED) obtained from UVOT/XRT afterglow. 
In section 2, we describe the discovery and follow-up observations of \thisgrb. The data acquisition and reduction procedures are explained in section 3. Section 4 presents the results obtained and a discussion on the possible classification of \thisgrb. Finally, we summarise the conclusions of this work in Section 5. We used the Hubble parameter $\rm H_{0}$ = 70 km $\rm sec^{-1}$ $\rm Mpc^{-1}$, density parameters $\rm \Omega_{\Lambda}= 0.73$, and $\rm \Omega_m= 0.27$ \citep{2011ApJS..192...14J}.

\begin{table*}
\caption{{\bf Characteristics of \thisgrb.} \tninty: Duration from \bat observations; Transient Duration: Duration of the transient repoerted by \gbm; HR: Ratio of the fluence in 25-50 \keV to  that in  15-25  \keV; \mvts : minimum variability time scale in 15-350 \keV;  \Ep: Peak energy as reported by \gbm team \citep{2021GCN.29536....1F}; $z$: photometric redshift obtained by fitting UVOT/X-ray SED; $E_{\rm \gamma, iso}$: Isotropic $\gamma$-ray energy calculated using the photometric redshift.}
\begin{center}

\begin{tabular}{ l l l }
\hline
\\
{ \bf Characteristics} & { \bf \thisgrb} & { \bf Detector} \\
\\
\hline
\\
\tninty~(15-350 \keV) & $3.76 \pm 0.26$ sec & \bat \\ \T
$T_{\rm 100}$~(25-294 \keV)  & 1.024 & \gbm \\ \T
HR  &   $1.40\pm0.02$ & \bat \\ \T
\mvts (sec) &  0.512 & \bat \\ \T
Spectral lag (ms) & 186$^{+68}_{-65}$  & \bat \\ \T
\Ep & $230$   & \gbm  \\ \T
Redshift $(z)$ & $0.55^{+0.90}_{-0.40}$ &  \xrt+UVOT\\ \T
$E_{\rm \gamma, iso}$ ($\rm erg$) & ($2.61\pm1.4) \times 10^{51}$ & \bat \\ 
\\

\hline
\end{tabular}
\end{center}
\label{tab:prompt_properties}
\end{table*} 

\begin{table}[!ht]
\caption{\tninty value of \thisgrb in different energy channels}
\centering
\large
\begin{tabular}{l  c  } 

 \hline
 \\
Energy range  &  \bf \tninty   \\
(\keV) & (sec) \\
\\
\hline 
\\
15 -- 350  & $3.76\pm0.26$ \\ \T
15 -- 25 & $4.17\pm0.12$ \\ \T
25 -- 50 & $3.78\pm0.14$ \\ \T
50 -- 100 & $3.49\pm0.72$ \\ \T
100 -- 350 & $3.79\pm0.26$ \\
\\
\hline
  \end{tabular}
\label{t90}  
\end{table}

\section{\thisgrb}
\label{thisgrb}
BAT on-board NASA's \swift space mission \citep{2005SSRv..120..143B} triggered \thisgrb on 17$^{th}$ February 2021 at 23:25:42 UT and provided the location of the source with coordinates RA, and DEC = 06h 30m 26s, +68d 42' 53" (J2000) respectively with an uncertainty of 3 arcmin \citep{2021GCN.29521....1S}. The \swift X-Ray Telescope (XRT) observations 97.8 seconds after the BAT trigger located an uncatalogued X-ray source at RA, and Dec = 06h 30m 20.82s, +68d 43' 29.9" (J2000) with an uncertainty of 3.7 arcseconds.
\swift Ultra-Violet and Optical telescope (UVOT) observations at 103 seconds after the BAT trigger detected a transient in white and U filters.

Several ground-based optical/NIR telescopes started observing the field of \thisgrb and reported the optical afterglow magnitudes\footnote{\url{https://gcn.gsfc.nasa.gov/other/210217A.gcn3}}. \gbm did not trigger the event automatically \citep{2021GCN.29536....1F}. Still, an automated, blind search for short GRBs (below the onboard triggering threshold) in \gbm identified a short GRB consistent with the \bat event in both time and location with a high significance value with a duration of 1.024 sec \citep{2021GCN.29536....1F}.

We also carried out the observations of the optical afterglow of \thisgrb with 1.3m DFOT located in ARIES at the earliest available opportunity \citep{2021GCN.29539....1K}. Once the afterglow crosses the limit of 1.3m DFOT, we carried out deep observations with the 3.6m DOT \citep{2021GCN.29591....1D}. The characteristics of the burst are presented in Table \ref{tab:prompt_properties}.

\begin{figure}[!ht]
\includegraphics[scale=0.28]{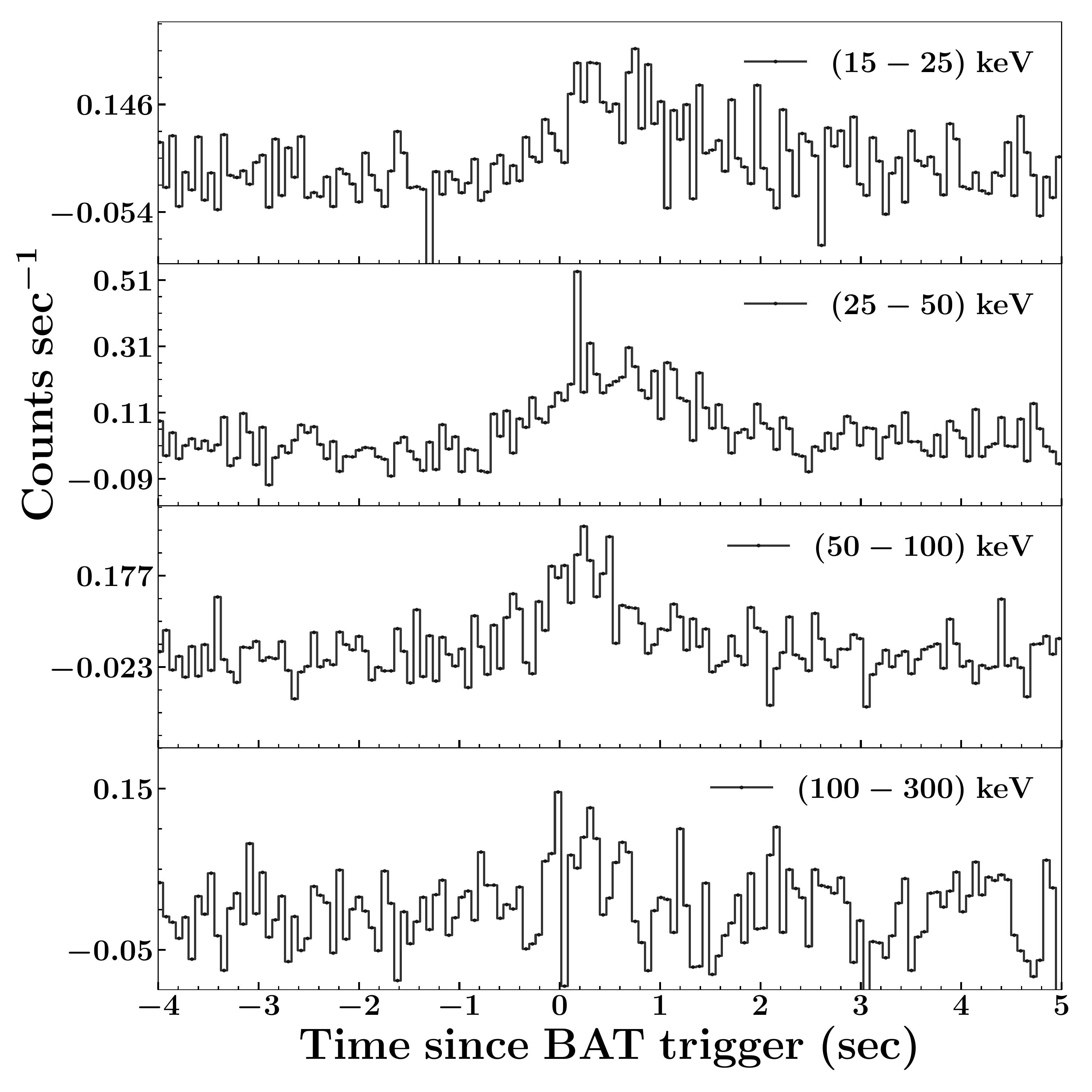}
\caption{ Background subtracted \bat count rate light curve of \thisgrb in different energy channels with a time resolution of 64 ms.}
\label{fig:promptlc}
\end{figure} 

\begin{table*}[!ht]
\caption{AB magnitudes of the afterglow of \thisgrb along with the magnitudes reported in the GCN circulars. Magnitudes are not corrected for Galactic extinction. All the upper limits are given with 3 sigma value.}
\centering
\begin{tabular}{c  c  c  c  c  c} 

 \hline
 \\
{\bf $\Delta$ t (days)}  &  \bf Filter  &\bf   Magnitude & \bf Telescope & \bf Reference \\
\\
\hline
\\
0.0021  & White & $18.80\pm0.06$ & {\it Swift}-UVOT & This Work\\ \T
2.2396  & White & $22.42\pm0.35$ & {\it Swift}-UVOT & This Work\\ \T
0.0048  & U & $18.75\pm0.16$ & {\it Swift}-UVOT & This Work\\ \T
0.75375 & R & $21.80\pm0.08$  & 1.3m DFOT  & This Work \\ \T
0.8220  & I & $  21.56\pm0.12$  & 1.3m DFOT  & This Work \\  \T
1.7010  & r & $  22.32\pm0.16$  & 3.6m DOT & This Work \\  \T
2.7401  & r & $  > 22.60$  & 3.6m DOT & This Work \\  \T
0.17791 & R & $20.80\pm0.30$  & KAIT & \cite{2021GCN.29533....1Z} \\  \T
0.0100  & z & $18.67\pm0.10$  & 2.0m Liverpool &  \cite{2021GCN.29535....1S}    \\  \T
0.00120 & r & $  18.60\pm0.20$  & 2.0m Liverpool & \cite{2021GCN.29535....1S}  \\ \T
0.0080 & i & $ 18.77\pm0.10$  & 2.0m Liverpool &  \cite{2021GCN.29535....1S}    \\   
\\
\hline
  \end{tabular}
\label{optical_data}  
\end{table*}

\begin{figure}[!ht]
\centering
\includegraphics[width=\columnwidth]{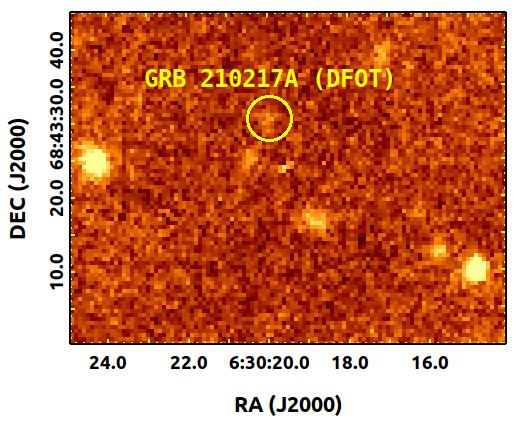}
\includegraphics[scale=0.34]{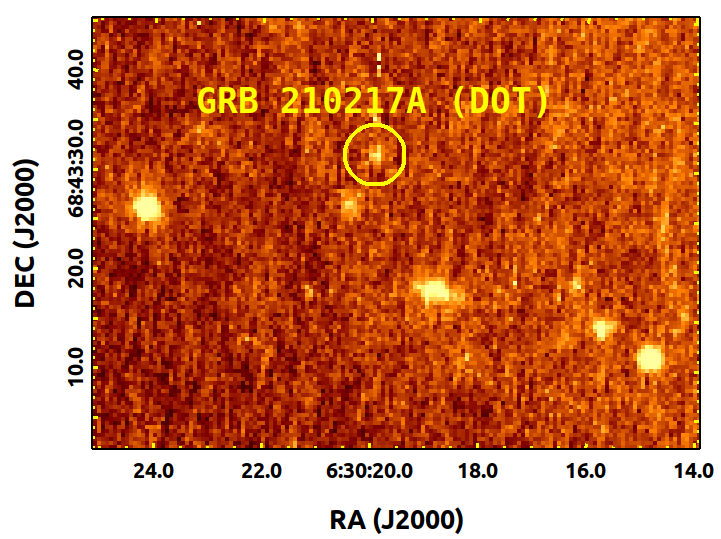}
\caption{The optical finding chart of GRB 210217A taken with 1.3m DFOT (left) and 3.6m DOT (right) telescopes. The afterglow is visible, and its position is marked with yellow circles.}
\label{fc}
\end{figure} 

\section{Data Acquisition and Reduction}
\label{data}
This section narrates the data acquisition and analysis from different space missions and ground-based telescopes as part of the present work. The \gbm data is not publicly available for \thisgrb as this burst was detected below the onboard triggering threshold\footnote{ \url{ https://gcn.gsfc.nasa.gov/fermi\_gbm\_subthresh\_archive.html}}. For the rest of the work we used the parameters provided by the \gbm team\footnote{ \url{https://gcn.gsfc.nasa.gov/fermi_gbm_subthreshold.html}}$^,$\footnote{\url{https://gcn.gsfc.nasa.gov/notices_gbm_sub/635297147.fermi}}. 

\subsection{\swift{\rm -BAT}}
\label{bat}

To extract the temporal and spectral properties of \thisgrb in high energy bands, the raw data of \bat (Observation Id: 01033264000) is obtained from the online portal of {\it Swift} Archive\footnote{\url{https://www.swift.ac.uk/swift_portal/}}. We reduced the data utilising the \sw{HEASOFT} (version-6.25).  The inbuilt pipelines \sw{batbinevt}, \sw{bathotpix}, and \sw{batmaskwtevt} are used to create detector plane image (DPI) followed by removal of hot pixels and mask weighting. Later, the mask-weighted BAT light curves are extracted using \sw{batbinevt} pipeline for different energy channels. 
The light curves in different energy channels are shown in Figure \ref{fig:promptlc}. Further, we estimated the \tninty duration of light curves in different energy channels and are tabulated in Table \ref{t90}. The \tninty value in the energy range 15-350 \keV is $3.76\pm0.26$ which is consistent with the value reported in \cite{2021GCN.29534....1S}.

To examine the spectral properties of the burst, we extracted the time-averaged BAT spectrum for the total duration of the burst starting from \swiftT$-0.065$  to \swiftT+0.489 sec (the start and end time of the pulse are identified using Bayesian binning of the light curve in the energy range 15-350 \keV).

We used the \sw{batbinevt} and  \sw{batdrmgen} pipelines to produce the spectrum and detector response matrix (DRM), respectively.
The resultant spectrum is fitted with the power-law and the cutoff power-law using the Multi-Mission Maximum Likelihood framework (\sw{3ML}) package \citep{2015arXiv150708343V}. Maximum likelihood estimation technique was used for choosing the best fit model. We got the maximum value for likelihood for a single power-law model with photon index ($\it \Gamma_{\rm BAT}$) of 1.99 $\pm$ 0.09. The fluence is $6.7 \pm 0.39 \times 10^{-7} \rm erg$ $\rm cm^{-2}$ in the 15-150 \keV band, which is consistent with the value reported by \citep{2021GCN.29534....1S}. 

\subsection{\swift{\rm -XRT \& UVOT}}
We obtained the X-ray flux light curve data in the 0.3-10 \keV energy band from \xrt Burst Analyser repository\footnote{\url{https://www.swift.ac.uk/}} hosted by the University of Leicester \citep{2009MNRAS.397.1177E}. The flux light curve is then converted to flux density at 5 \keV using the relations given by \cite{Gehrels:2008ApJ} for further analysis.

\begin{equation}
    F_\nu,x=4.13\times 10^{11} \frac{F_x(2-\Gamma_x)E_0^{1-\Gamma_x}}{E_2^{2-\Gamma_x}-E_1^{2-\Gamma_x}}.
\end{equation}
 
\noindent
where, $E_1$ and $E_2$ are lower and upper bounds of band pass in keV, $E_0$ is the energy in \keV at which flux density is calculated, $ \Gamma_x$ is the X-ray photon index. $F_x$ is the measured flux in erg cm$^{-2}$ sec$^{-1}$. 

We downloaded the \uvot data from the online \swift data archive page\footnote{\url{http://swift.gsfc.nasa.gov/docs/swift/archive/}}. We analyzed the UVOT data using standard pipeline \sw{uvotproduct} of \sw{heasoft} software version 6.25 with the latest calibration database. A source region of 5 arcsec and a background region of 25 arcsec aperture radius are extracted for the photometric analysis of the burst. 
We detected a source in white and U filter. Table \ref{optical_data} shows the magnitudes of the source in these filters.

\subsection{1.3m DFOT and 3.6m DOT}

We started observations with 1.3 DFOT on 2021-02-18 at 17:31:22 UT located at Devasthal observatory of ARIES, India, for the follow-up observations of the optical afterglow of \thisgrb. We observed a set of $30$ images with an exposure of 120s each in the Bessel R filter and a set of 20 images (120s exposure) in the Bessel I filter. We detected the optical afterglow of \thisgrb in the stacked frames within the {\it Swift} XRT error circle (left panel of Figure \ref{fc}). Later, for deep observations, we observed the field with Aries Devasthal Faint Object Spectrograph and Camera (ADFOSC) mounted on the 3.6m DOT around $\sim 1.7$ days after the burst. We took four consecutive images in r-band with an exposure of 900 sec each. We corrected the science images acquired from these telescopes for bias, flat and cosmic rays using \sw{Astropy} module of \sw{Python}.

\begin{figure}
\centering
\includegraphics[scale=0.35]{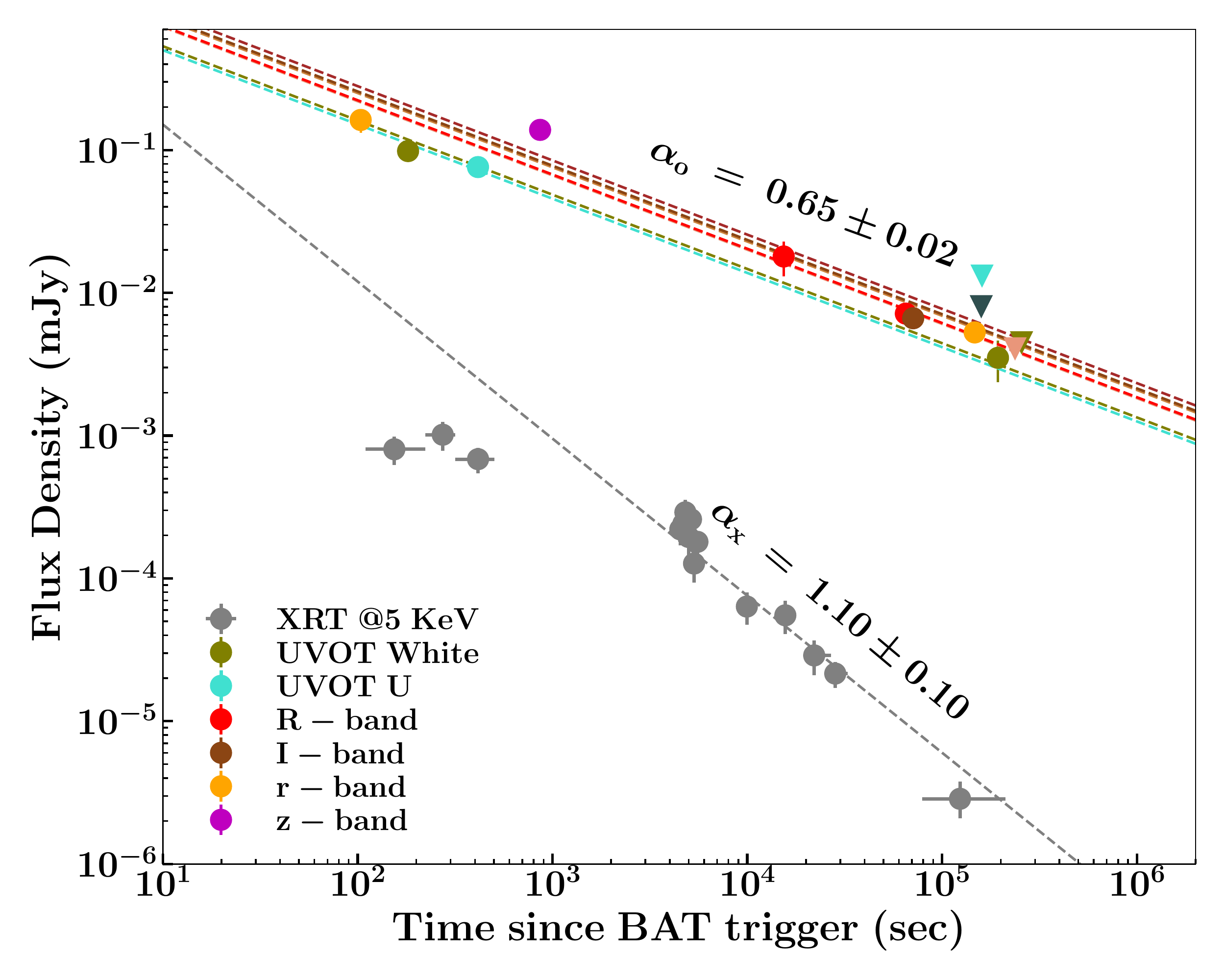}
\caption{Multiwavelength afterglow light curves of \thisgrb. Optical data points are corrected for galactic extinction. The X-ray data (shown in grey) are taken from \swift-XRT page. The best fit models are also shown with dashed lines.}
\label{lc}
\end{figure}

We stacked the images to enhance the signal-to-noise ratio. The source is visible in stacked image \citep{2021GCN.29591....1D}. The finding chart for the same is shown in the right panel of Figure \ref{fc}. We used the \sw{DAOPHOT} package to perform point spread function (PSF) photometry which was calibrated against the \sw{Panstarrs} catalog resulting in apparent magnitudes listed in Table \ref{optical_data}. The magnitudes are converted to flux density after correcting for galactic extinction.  Figure \ref{lc} shows the multi-band light curve of \thisgrb constructed using our data and those reported in GCN circulars. 

Further, we fitted the X-ray and optical R band light curves of GRB 210217A using a single power law of the form $F=F_{0}t^{-\alpha}$, where t corresponds to time, and $\alpha$ is the decay index. Due to the limited number of data points in other optical bands, it is not possible to fit these. Therefore, We overplotted the light curves in other bands using the R band decay index. The light curves with power laws are shown in Figure \ref{lc}.

\section{Results and Discussions}
\label{results}
This section presents the results obtained from the multi-band analysis and possible classification scenario of GRB 210217A. 

\subsection{ {Photometric Redshift}}
\label{photoz}
\begin{figure}[!ht]
\centering
\includegraphics[scale=0.33]{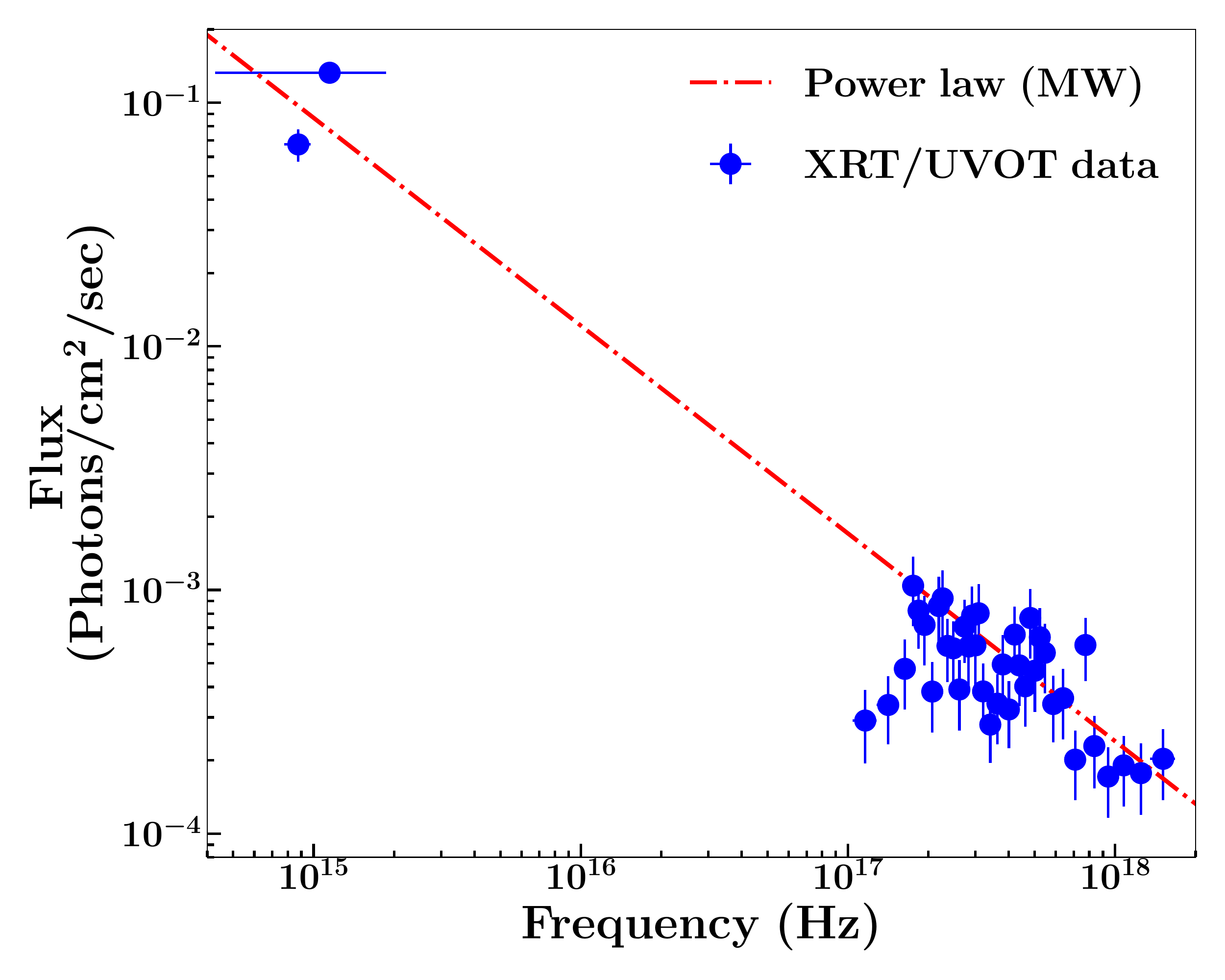}
\caption{The SED obtained using joint \swift UVOT and XRT afterglow data for \thisgrb. Data points are shown using blue circles. The SED is best fitted with a power-law model (shown with red dashed dotted line) with Milky way extinction law.}
\label{sed}
\end{figure}

To estimate the photometric redshift of \thisgrb, we analyzed the joint XRT and UVOT SED in the time interval of $\sim$ 100-200 sec. We didn't observe any spectral evolution during this time interval. Using the XRT data and UVOT magnitudes, we created the SED  
following the methodology described in \cite{2020ApJ...898...42C, 2021MNRAS.505.4086G}. We fitted the SED using power-law and broken power-law models using \sw{XSPEC} \citep{1996ASPC..101...17A}. The Galactic and intrinsic absorber components (\sw{phabs} and \sw{zphabs}) are also included from XSPEC models . The Galactic absorption is fixed to $\rm NH_{\rm Gal}=8.69 \times 10^{20}{\rm cm}^{-2}$ \citep{2013MNRAS.431..394W}. We further included two dust components using the XSPEC model \sw{zdust}, one at redshift $z$= 0 for galactic dust component, and the other for the intrinsic dust component with varying redshift which provided the redshift information. The Galactic reddening is fixed at 0.0847 (E(B-V)) conforming to the map of \cite{sch11}. The SED is fitted with Milky Way, Large and Small Magellanic Clouds (MW, LMC, and SMC) extinction laws \citep{pei92} at the redshift of the burst. Although all these models well explain the SED, a minimum $\chi^{2}$ value is obtained for Milky way extinction law with a power-law model. The value of the spectral index is 1.85 $\pm$  0.13, and that of photometric redshift is $0.55^{+0.9}_{-0.4}$ for \thisgrb. The SED is shown in Figure \ref{sed}.

\subsection{Hardness Ratio and Peak Energy}

\begin{figure*}[!ht]
\includegraphics[scale=0.36]{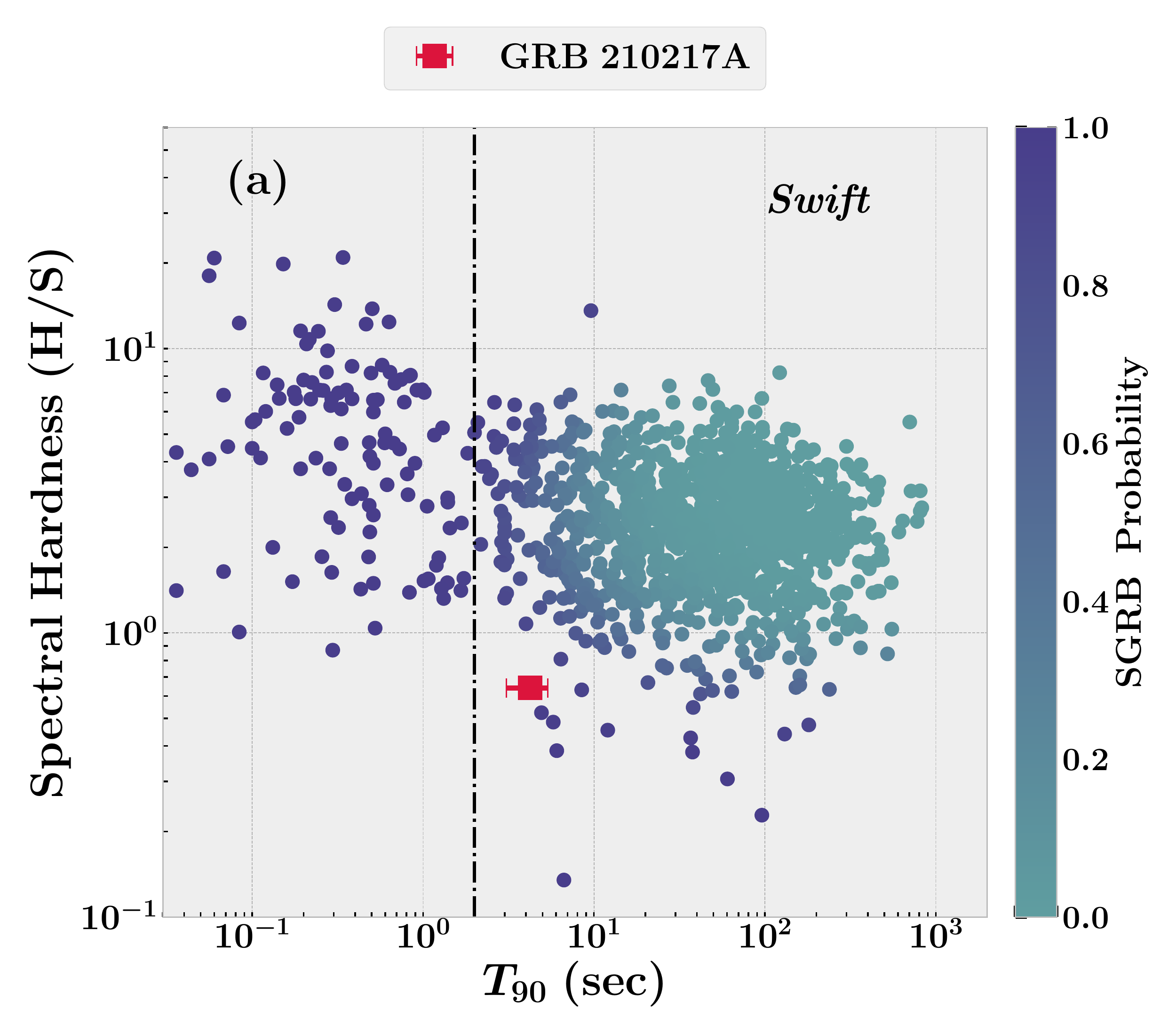}
\includegraphics[scale=0.36]{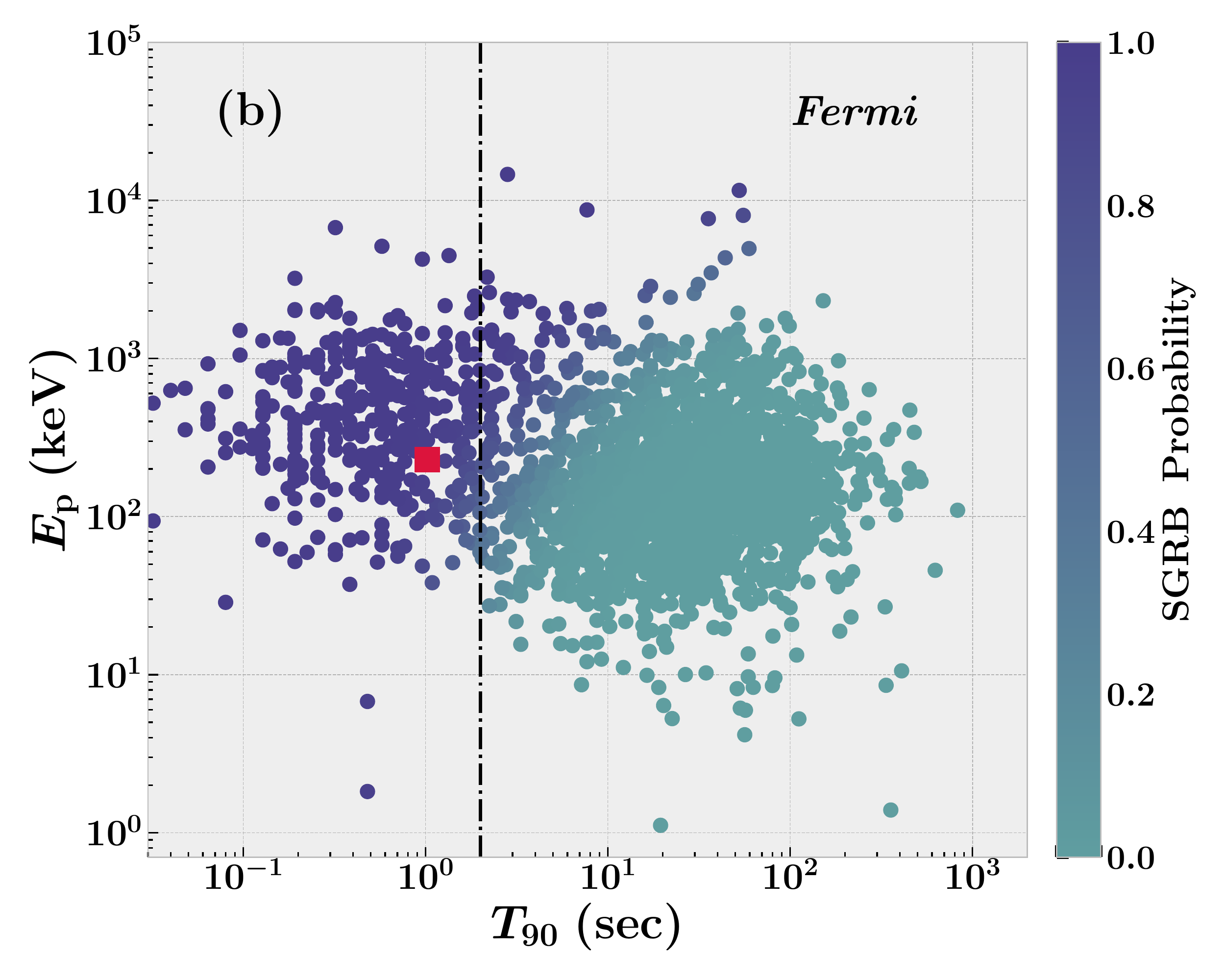}
\includegraphics[scale=0.36]{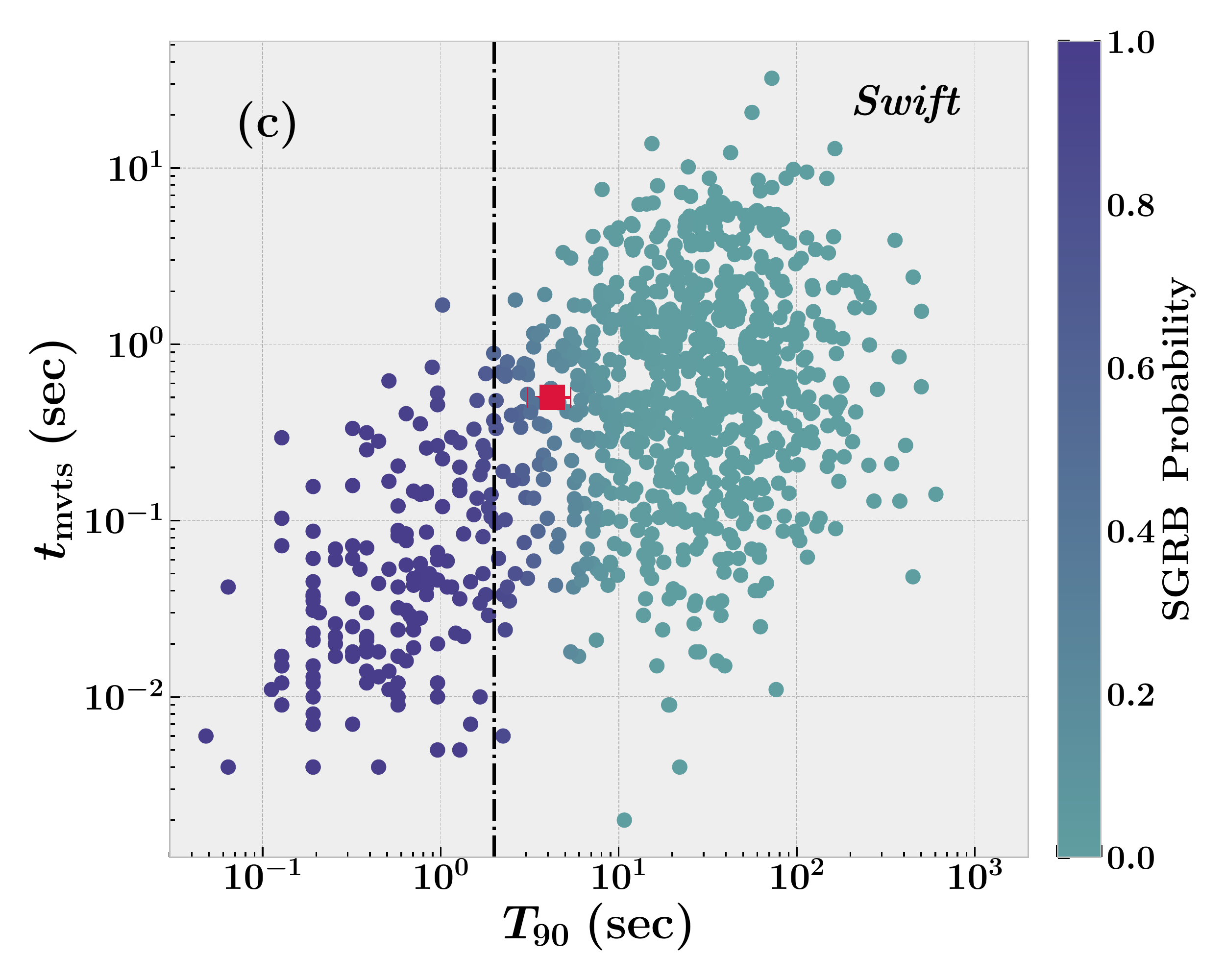}
\includegraphics[scale=0.34]{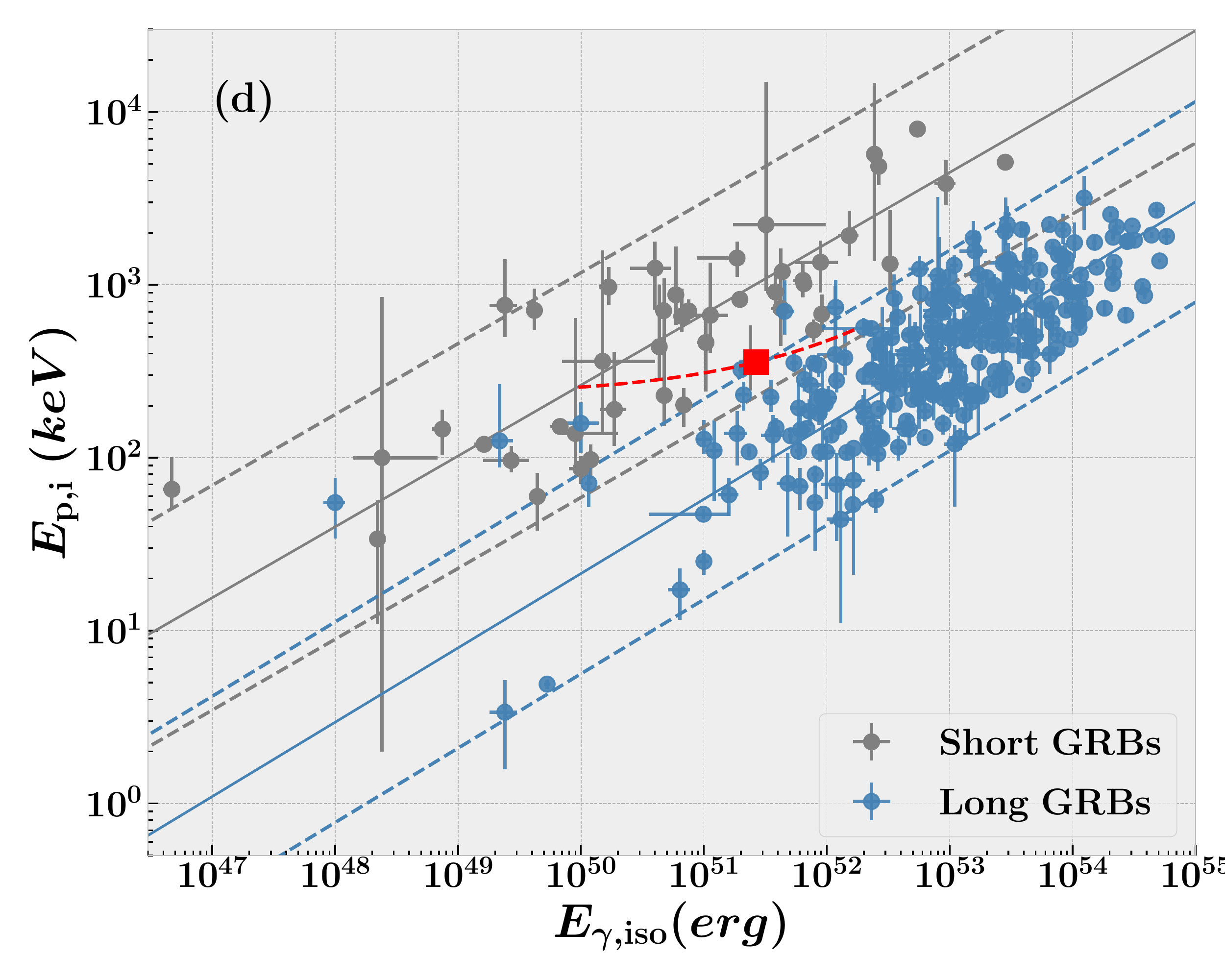}
\caption{(a) The spectral hardness and \tninty for \thisgrb (shown with a red square) along with the GRB sample (for short and long GRBs) presented in \cite{2017ApJ...848L..14G}. 
(b) \thisgrb in \Ep - \tninty plane with the dataset available from \fermi- GBM catalog.
(c) Minimum variability time scale (\mvts) as a function of \tninty for \thisgrb along with the short and long GRBs sample studied by \cite{2015ApJ...811...93G}. 
The colour scale on the right side of the Figure shows the probability of the GRB being a short GRB and the vertical dashed line at 2 sec shows the boundary separating the short and long classes of GRB. 
(d) \thisgrb in the Amati plane along with the data points for long  and short GRBs taken from \cite{min20b}. The solid lines shows the best correlation between $E_{\rm p,i}-E_{\rm \gamma, iso}$along with 2-sigma correlation presented by dashed line of respective colors for short (grey) and long class (steelblue). $E_{\rm p,i}-E_{\rm \gamma, iso}$
 for \thisgrb is shown with a red square. The red dashed line show the $E_{\rm p,i}-E_{\rm \gamma, iso}$ for the redshift range considering the uncertainty in photometric redshift.  }
\label{classification}
\end{figure*} 
We estimated the HR using fluence ratio in two different energy channels: 15-25 \keV and 50-100 \keV energy bands of \bat and used it for the classification of \thisgrb. The HR during \tninty is found to be 0.64 $\pm$ 0.005. We collected the sample of GRBs from \bat catalog\footnote{\url{https://swift.gsfc.nasa.gov/results/batgrbcat/summary_cflux/summary_T100/summary_pow_energy_fluence.txt }} and estimated the hardness ratio using the fluence in the same energy window as used for \thisgrb. We fitted this sample with the Bayesian Gaussian mixture model (BGMM), which is a machine learning algorithm supported by scikit-learn \citep{scikit-learn}. Using this algorithm, we estimated the probability of \thisgrb being a short GRB as 98.2\%.
The probabilities of the whole sample being a short GRB along with \thisgrb are shown in Figure \ref{classification} (a).

As \swift BAT has narrow spectral coverage, we could not determine \Ep using spectral fitting. Therefore, we used the $E_{p}$, peak energy and \tninty values provided by \gbm along with the $E_{p}$-\tninty values from the \gbm catalog to identify the class of \thisgrb. We again fitted the distribution with BGMM and found a probability of 96\% for \thisgrb being a short GRB. The probability map is shown in Figure \ref{classification} (b).

\subsection{Minimum variability time scale}

High energy light curves of GRBs are highly variable and can be explained with the GRB central engine resulting from internal shocks. Hence, minimum variability time scale \citep[][\mvts]{2013MNRAS.432..857M} gives an idea about the central engine, the source emission radius ($\rm R_{\rm c}$), and  the minimum Lorentz factor \citep[][\lmin]{2015ApJ...805...86S} of GRBs. In general, the \mvts value for long bursts is larger with an average value of 200 ms than their short counterparts having a mean value of 10 ms, indicating that short GRBs have a more compact central engine \citep{MacLachlan2012,2013MNRAS.432..857M}. We measured the \mvts for \thisgrb using the method described in \cite{2015ApJ...811...93G}. The estimated value of \mvts is $\sim$ 0.512 sec for \thisgrb, which we further used for its classification. We collected \mvts values for short and long GRB samples from \cite{2015ApJ...811...93G}. Using BGMM, we found the probability of \thisgrb being a short GRB equal to 28\% (Figure \ref{classification} (c)).

\subsection{Spectral lag}

We calculated the spectral lag using bat light curves in energy channels; 15-25 \keV and 50-100 \keV. We estimated the using the cross-correlation function (CCF) and uncertainties in the CCF following the method described in \cite{2015MNRAS.446.1129B}. We fitted the CCF with an asymmetric Gaussian function using emcee \citep{Foreman_Mackey_2013} to find its global maximum, which represents the lag in two light curves. We found a positive lag 86$^{+68}_{-65}$ ms for the burst. The average value of lag for short and long GRBs are $16.5\pm7.5$ and $375.1\pm69.6$ ms, respectively \citep{2015MNRAS.446.1129B}. The value is lying more close to the mean value of short GRBs within the errorbar. With the intermediate value of spectral lag lying between the mean value of two classes, it is hard to classify the burst.

\subsection{Amati correlation}

We also used the well-known correlation between the isotropic equivalent energy emitted in the high energy regime, $ E_{\rm \gamma, iso}$ and \Ep (the energy at which $ \nu F_{\nu} $ is maximum) to classify \thisgrb\citep{Amati_2002, Amati_2006, min20b}. A power-law well fits this correlation with an index of $ a \simeq 0.4 $; however, the correlation regions are well separated for different classes \citep{min20}. 

For the redshift range (0.15-1.45) obtained from SED (section \ref{photoz}), we estimated the $ E_{\rm \gamma, iso}$ using the following equation \ref{eiso}

\begin{center}
\begin{equation}
\label{eiso}
    E_{\gamma,iso}=K_{bol}\times \frac{4 \pi d_{L}^2}{1+z}f_\gamma.
\end{equation}
\end{center} 

where, $K_{bol}$ is the bolometric correction factor, $d_L$ is the luminosity distance and $f_\gamma$ is the fluence (erg $cm^{-2}$). For \bat energy range (15-150 \keV) we used $K_{bol}$ = 5 \citep{fong2015} and the fluence value calculated in section \ref{bat}.  We also estimated the $E_{\rm p,i}$ (peak energy in the source frame) for the estimated redshift range. As \bat covers only a small energy range, we used the \Ep and burst duration as reported by \gbm team \citep{2021GCN.29536....1F}. The red dashed line in the Figure \ref{classification} (d) shows the $ E_{\rm \gamma, iso}$ - $E_{\rm p,i}$ for the redshift range of \thisgrb. \thisgrb lies in the region in between long and short GRBs.

\section{Summary}
\label{summary}
We report a detailed analysis of \thisgrb using publicly available multi-wavelength observations, including our observations from our ARIES telescopes. We try to find the true class of \thisgrb using various methods described in the literature. We calculated the photometric redshift using Swift-XRT/UVOT data. Using this estimated redshift, we calculated the isotropic equivalent energy and peak energy in the source frame. \thisgrb lies at the boundary between short and long class in the Amati plane.

We also calculated the HR, minimum variability timescale, spectral lag and fitted the \tninty-HR, \tninty-Ep, \tninty-mvts distribution using BGMM. We found a probability of \thisgrb being a short GRB equal to 98.2\%, 96\%, and 28\% in these cases, respectively. It is hard to conclude if \thisgrb belongs to the long or the short class as some of the properties belong to long GRBs and others to short. 

The host studies can clarify the true class of this burst. The fact that two categories have different progenitors, they are originated in different kinds of host galaxies. Long GRBs are located in the star-forming young population of galaxies; the short GRBs belongs to the old population of galaxies \citep{Li_2016}. Long GRBs are generally located in bright star-forming regions with minimal offsets from the centre of the host galaxy on a galactic scale. In contrast, due to significant merger scale times, short GRBs have large offsets from the centre of their galaxies \citep{2002AJ....123.1111B, Fong_2013}. The host observations are not available for \thisgrb. However, the host observations can give a clue about the class of \thisgrb. 
%We will plan to observe the host of \thisgrb using 3.6m DOT. 

\section*{Acknowledgements}

KM, RG, and SBP acknowledge BRICS grant {DST/IMRCD/BRICS/PilotCall1/ProFCheap/2017(G)} for the financial support. KGA is partially supported by the Swarnajayanti Fellowship Grant No.DST/SJF/PSA-01/2017-18, MATRICS grant MTR/2020/000177 of SERB, and a grant from the Infosys Foundation. This research is based on observations obtained at the 3.6m Devasthal Optical Telescope (DOT) during observing cycles DOT-2021-C1, a National Facility runs and managed by Aryabhatta Research Institute Observational Sciences (ARIES), an autonomous Institute under the Department of Science and Technology, Government of India. This research has used data obtained from the High Energy Astrophysics Science Archive Research Center (HEASARC) and the Leicester Database and Archive Service (LEDAS), provided by NASA's Goddard Space Flight Flight Flight Center and the Department of Physics and Astronomy, Leicester University, UK, respectively.

\section{Appendix}
\subsection{Afterglow Properties}
\begin{table*}
\centering
%\scriptsize

\caption{Optical and X-ray spectral indices obtained from fitting SEDs at different epochs and their best describe spectral regime. $p$ is the mean value of the electron distribution indices calculated using temporal and spectral indices in that spectral regime}
\begin{tabular}{c c c c}
\hline
\textbf{Time interval (s)} & \textbf{$\bf \beta_{\rm \bf opt}$} & \textbf{$\bf \beta_{\rm \bf x-ray}$} & \textbf{\begin{tabular}[c]{@{}c@{}}$p$\\ (Spectral regime)\end{tabular}} \\
\hline
\small
0.5-5 $\times$ $10^{2}$ & 0.61$^{+0.03}_{-0.02}$ & 0.89$^{+0.32}_{-0.30}$ & \begin{tabular}[c]{@{}c@{}}2.00 $\pm$ 0.35\\( $\nu_{\rm opt}$  $<$ $\nu_{\rm x-ray}$  $<$  $\nu_{\rm c}$)\end{tabular} \\ \hline
6.5-19.3 $\times$ $10^{4}$ & 0.65$^{+0.08}_{-0.07}$ & 1.32$^{+0.28}_{-0.21}$ & \begin{tabular}[c]{@{}c@{}}2.43 $\pm$ 0.36\\ ($\nu_{\rm opt}$ $<$ $\nu_{\rm c}$ $<$ $\nu_{\rm x-ray}$)\end{tabular} \\ \hline

\end{tabular}
\label{tab:optical_SED}
\end{table*}

\begin{figure*}[!ht]
\centering
\includegraphics[scale=0.45]{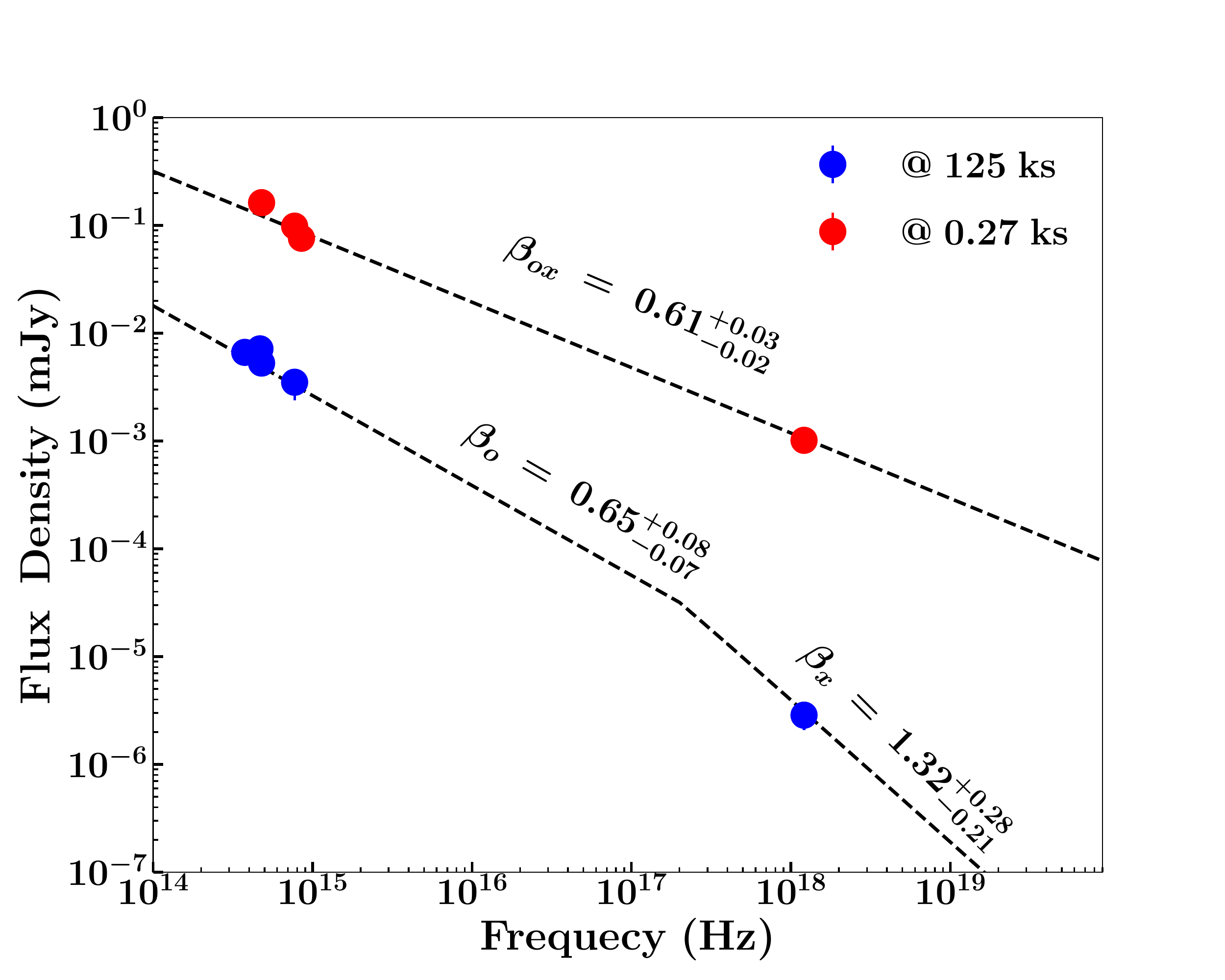}
\caption{The optical/X-ray SEDs at two different epochs ($\sim $ 270 sec and at $\sim $ 1.45 days after the trigger). The SED at the earlier epoch is well fitted with a simple power-law with a spectral index of 0.61$^{+0.03}_{-0.02}$. However, a spectral break can be seen at the later epoch which is identified as the synchrotron cooling break.}
\label{opt_sed}
\end{figure*}

The afterglow of the GRB can be well explained by the synchrotron fireball model \citep{1999PhR...314..575P}. The spectra, as well as the light curves, consist of a combination of power-law and broken power-law characterized by electron distribution index $p$ \citep{Piran_2005, Sari_1998}. We used the spectral and temporal indices to constrain $p$ and the break frequencies using well-known closure relations \citep{Sari_1998}. 
For this purpose, we fitted the X-ray and optical light curves/spectra at different epochs with single and broken power-law models. Both X-ray and optical light curves are well explained with a single power with indices value of $1.10 \pm 0.1$ and $0.65\pm0.02$, respectively. 
The values of spectral indices at different epochs are given in Table \ref{tab:optical_SED}. 
At around 0.3 ks, the spectral index is almost the same for optical and X-ray within the errorbar, suggesting no cooling break between X-ray to optical data.  
However, we found that the X-ray spectral index is almost two times the optical index at later epochs, indicating some break. 
So, we further created the spectral energy distributions using optical and X-ray data at two epochs centred at $\sim 0.27 $ ks and $\sim 125 $ ks. We fitted the SEDs with power-law and broken power-law models. The SED at the early epoch is best fitted with a single power-law with an index of $0.607\pm0.02$. However, the SED at the later epoch is best fitted with a broken power-law with indices $0.65^{+0.08}_{-0.07}$ (pre-break) and $1.32^{+0.28}_{-0.21}$ (post-break) and a break at  $1.7\pm0.3 \times 10^{17}$ Hz, which we identify as a cooling break. The best-fitted SEDs are shown in Figure \ref{opt_sed}.

\vspace{-1em}


\begin{thebibliography}{}
\bibitem[Ahumada et al.(2021)]{2021NatAs.tmp..142A} Ahumada, T., Singer, L.~P., Anand, S., et al.\ 2021, Nature Astronomy. doi:10.1038/s41550-021-01428-7
\bibitem[Amati et al.(2002)]{Amati_2002} Amati, L., Frontera, F., Tavani, M., et al.\ 2002, \aap, 390, 81. doi:10.1051/0004-6361:20020722
\bibitem[Amati(2006)]{Amati_2006} Amati, L.\ 2006, \mnras, 372, 233. doi:10.1111/j.1365-2966.2006.10840.x
\bibitem[Antonelli et al.(2009)]{2009A&A...507L..45A} Antonelli, L.~A., D'Avanzo, P., Perna, R., et al.\ 2009, \aap, 507, L45. doi:10.1051/0004-6361/200913062
\bibitem[Arnaud(1996)]{1996ASPC..101...17A} Arnaud, K.~A.\ 1996, Astronomical Data Analysis Software and Systems V, 101, 17
\bibitem[Barthelmy et al.(2005)]{2005SSRv..120..143B} Barthelmy, S.~D., Barbier, L.~M., Cummings, J.~R., et al.\ 2005, \ssr, 120, 143. doi:10.1007/s11214-005-5096-3
\bibitem[Bernardini et al.(2015)]{2015MNRAS.446.1129B} Bernardini, M.~G., Ghirlanda, G., Campana, S., et al.\ 2015, \mnras, 446, 1129. doi:10.1093/mnras/stu2153
\bibitem[Bloom et al.(2002)]{2002AJ....123.1111B} Bloom, J.~S., Kulkarni, S.~R., \& Djorgovski, S.~G.\ 2002, \aj, 123, 1111. doi:10.1086/338893
\bibitem[Breeveld et al.(2011)]{bre11} Breeveld, A.~A., Landsman, W., Holland, S.~T., et al.\ 2011, Gamma Ray Bursts 2010, 1358, 373. doi:10.1063/1.3621807
%\bibitem[Burrows et al.(2021)]{2021GCN.29538....1B} Burrows, D.~N., Evans, P.~A., Osborne, J.~P., et al.\ 2021, GRB Coordinates Network, Circular Service, No. 29538, 29538
\bibitem[Chand et al.(2020)]{2020ApJ...898...42C} Chand, V., Banerjee, A., Gupta, R., et al.\ 2020, \apj, 898, 42. doi:10.3847/1538-4357/ab9606
\bibitem[Cheng et al.(1995)]{1995A&A...300..746C} Cheng, L.~X., Ma, Y.~Q., Cheng, K.~S., et al.\ 1995, \aap, 300, 746
\bibitem[Dimple et al.(2021)]{2021GCN.29591....1D} Dimple, Misra, K., Ghosh, A., et al.\ 2021, GRB Coordinates Network, Circular Service, No. 29591, 29591
\bibitem[Evans et al.(2009)]{2009MNRAS.397.1177E} Evans, P.~A., Beardmore, A.~P., Page, K.~L., et al.\ 2009, \mnras, 397, 1177. doi:10.1111/j.1365-2966.2009.14913.x
%\bibitem[Evans et al.(2021)]{2021GCN.29525....1E} Evans, P.~A., Goad, M.~R., Osborne, J.~P., et al.\ 2021, GRB Coordinates Network, Circular Service, No. 29525, 29525
\bibitem[Fenimore et al.(1995)]{1995ApJ...448L.101F} Fenimore, E.~E., in 't Zand, J.~J.~M., Norris, J.~P., et al.\ 1995, \apjl, 448, L101. doi:10.1086/309603
\bibitem[Fishman \& Meegan(1995)]{1995ARA&A..33..415F} Fishman, G.~J. \& Meegan, C.~ A.\ 
1995, ARAA, 33, 415. doi:10.1146/annurev.aa.33.090195.002215
\bibitem[Fletcher \& Fermi-GBM Team(2021)]{2021GCN.29536....1F} Fletcher, C. \& Fermi-GBM Team\ 2021, GRB Coordinates Network, Circular Service, No. 29536, 29536
\bibitem[Fong et al.(2013)]{Fong_2013} Fong, W., Berger, E. and Chornock, R.et al. \ 2013, \apj, 769, 1.
\bibitem[Fong et al. (2015)]{fong2015}Fong, W. and Berger, E. and Margutti, R. et al.\ 2015, \apj, 102. doi:10.1088/0004-637x/815/2/102
\bibitem[Foreman-Mackey et al.(2013)]{Foreman_Mackey_2013} Foreman-Mackey, D., Hogg, D.~W., Lang, D., et al.\ 2013, pasp, 125, 306. doi:10.1086/670067
\bibitem[Gal-Yam et al.(2006)]{2006Natur.444.1053G} Gal-Yam, A., Fox, D.~B., Price, P.~A., et al.\ 2006, \nat, 444, 1053. doi:10.1038/nature05373
\bibitem[Gehrels et al. (2008)]{Gehrels:2008ApJ} Gehrels, N., Barthelmy, S.~D. and Burrows, D.~N. et al. \ 2008, \apj, 689, 2. doi: 10.1086/592766.
\bibitem[Ghirlanda et al.(2009)]{2009A&A...496..585G} Ghirlanda, G., Nava, L., Ghisellini, G., et al.\ 2009, \aap, 496, 585. doi:10.1051/0004-6361/200811209
\bibitem[Goldstein et al.(2017)]{2017ApJ...848L..14G} Goldstein, A., Veres, P., Burns, E., et al.\ 2017, \apjl, 848, L14. doi:10.3847/2041-8213/aa8f41
\bibitem[Golkhou et al.(2015)]{2015ApJ...811...93G} Golkhou, V.~Z., Butler, N.~R., \& Littlejohns, O.~M.\ 2015, \apj, 811, 93. doi:10.1088/0004-637X/811/2/93
\bibitem[Gupta et al.(2021)]{2021MNRAS.505.4086G} Gupta, R., Oates, S.~R., Pandey, S.~B., et al.\ 2021, \mnras, 505, 4086. doi:10.1093/mnras/stab1573
\bibitem[Jarosik et al.(2011)]{2011ApJS..192...14J} Jarosik, N., Bennett, C.~L., Dunkley, J., et al.\ 2011, \apjs, 192, 14. doi:10.1088/0067-0049/192/2/14
%\bibitem[Kalberla et al.(2005)]{kal05} Kalberla, P.~M.~W., Burton, W.~B., Hartmann, D., et al.\ 2005, \aap, 440, 775. doi:10.1051/0004-6361:20041864
\bibitem[Kaneko et al.(2015)]{2015MNRAS.452..824K} Kaneko, Y., Bostanc{\i}, Z.~F., G{\"o}{\u{g}}{\"u}{\c{s}}, E., et al.\ 2015, \mnras, 452, 824. doi:10.1093/mnras/stv1286
\bibitem[Kouveliotou et al.(1993)]{1993ApJ...413L.101K} Kouveliotou, C., Meegan, C.~A., Fishman, G.~J., et al.\ 1993, \apjl, 413, L101. doi:10.1086/186969
\bibitem[Qin \& Chen(2013)]{2013MNRAS.430..163Q} Qin, Y.-P. \& Chen, Z.-F.\ 2013, \mnras, 430, 163. doi:10.1093/mnras/sts547
\bibitem[Kumar et al.(2021)]{2021GCN.29539....1K} Kumar, A., Gupta, R., Ghosh, A., et al.\ 2021, GRB Coordinates Network, Circular Service, No. 29539, 29539
\bibitem[Li et al.(2016)]{Li_2016} Li, Y., Zhang, B., \& L{\"u}, H.-J.\ 2016, \apjs, 227, 7. doi:10.3847/0067-0049/227/1/7
\bibitem[MacLachlan et al.(2012)]{MacLachlan2012} MacLachlan, G. A., Shenoy, A., Sonbas, E. et al. \ 2012, \mnras, 425, L32-L35. doi: 10.1111/j.1745-3933.2012.01295.x
\bibitem[MacLachlan et al.(2013)]{2013MNRAS.432..857M} MacLachlan, G.~A., Shenoy, A., Sonbas, E., et al.\ 2013, \mnras, 432, 857. doi:10.1093/mnras/stt241
\bibitem[Minaev \& Pozanenko(2020)]{min20} Minaev, P.~Y. \& Pozanenko, A.~S.\ 2020, \mnras, 492, 1919. doi:10.1093/mnras/stz3611
\bibitem[Minaev \& Pozanenko(2020)]{min20b} Minaev, P.~Y. \& Pozanenko, A.~S.\ 2020, Astronomy Letters, 46, 573. doi:10.1134/S1063773720090042
\bibitem[Norris(2002)]{2002ApJ...579..386N} Norris, J.~P.\ 2002, \apj, 579, 386. doi:10.1086/342747
%\bibitem[Pankov et al.(2021)]{2021GCN.29577....1P} Pankov, N., Belkin, S., Pozanenko, A., et al.\ 2021, GRB Coordinates Network, Circular Service, No. 29577, 29577
\bibitem[Pedregosa et al. (2011)]{scikit-learn} Pedregosa, F. and Varoquaux, G. et al.\ 2011, JMLR, 12, 2825--2830.
\bibitem[Pei(1992)]{pei92} Pei, Y.~C.\ 1992, \apj, 395, 130. doi:10.1086/171637
\bibitem[Piran(1999)]{1999PhR...314..575P} Piran, T.\ 1999, \physrep, 314, 575. doi:10.1016/S0370-1573(98)00127-6
\bibitem[Piran(2004)]{Piran_2005} Piran, T.\ 2004, Reviews of Modern Physics, 76, 1143. doi:10.1103/RevModPhys.76.1143
\bibitem[Qin et al.(2000)]{2000PASJ...52..759Q} Qin, Y.-P., Xie, G.-Z., Xue, S.-J., et al.\ 2000, PASJ, 52, 759. doi:10.1093/pasj/52.5.759
\bibitem[Sakamoto et al.(2021)]{2021GCN.29534....1S} Sakamoto, T., Barthelmy, S.~D., Cummings, J.~R., et al.\ 2021, GRB Coordinates Network, Circular Service, No. 29534, 29534
\bibitem[Sari et al.(1998)]{Sari_1998} Sari, R., Piran, T., \& Narayan, R.\ 1998, \apjl, 497, L17. doi:10.1086/311269
%\bibitem[Schady et al.(2007)]{schady07} Schady, P., Mason, K.~O., Page, M.~J., et al.\ 2007, \mnras, 377, 273. doi:10.1111/j.1365-2966.2007.11592.x
\bibitem[Schlafly \& Finkbeiner(2011)]{sch11} Schlafly, E.~F. \& Finkbeiner, D.~P.\ 2011, \apj, 737, 103. doi:10.1088/0004-637X/737/2/103
\bibitem[Shrestha et al.(2021)]{2021GCN.29535....1S} Shrestha, M., Smith, R., Melandri, A., et al.\ 2021, GRB Coordinates Network, Circular Service, No. 29535, 29535
\bibitem[Simpson et al.(2021)]{2021GCN.29521....1S} Simpson, K.~K., Gropp, J.~D., Kennea, J.~A., et al.\ 2021, GRB Coordinates Network, Circular Service, No. 29521, 29521
%\bibitem[Simpson \& Swift/UVOT Team(2021)]{2021GCN.29555....1S} Simpson, K.~K. \& Swift/UVOT Team\ 2021, GRB Coordinates Network, Circular Service, No. 29555, 29555
\bibitem[Sonbas et al.(2015)]{2015ApJ...805...86S} Sonbas, E., MacLachlan, G.~A., Dhuga, K.~S., et al.\ 2015, \apj, 805, 86. doi:10.1088/0004-637X/805/2/86
\bibitem[Tavani(1998)]{1998ApJ...497L..21T} Tavani, M.\ 1998, \apjl, 497, L21. doi:10.1086/311276
\bibitem[Vianello et al.(2015)]{2015arXiv150708343V} Vianello, G., Lauer, R.~J., Younk, P., et al.\ 2015, arXiv:1507.08343
\bibitem[Willingale et al. (2013)]{2013MNRAS.431..394W} Willingale, R., Starling, R.~L.~C. and Beardmore, A.~P., et al.\ 2013, \mnras, 431, 394-404. doi: 10.1093/mnras/stt175
\bibitem[Yi et al.(2006)]{2006MNRAS.367.1751Y} Yi, T., Liang, E., Qin, Y., et al.\ 2006, \mnras, 367, 1751. doi:10.1111/j.1365-2966.2006.10083.x
\bibitem[Zheng et al.(2021)]{2021GCN.29533....1Z} Zheng, W., Filippenko, A.~V., \& KAIT GRB Team\ 2021, GRB Coordinates Network, Circular Service, No. 29533, 29533
%\bibitem[Zhu et al.(2021)]{2021GCN.29523....1Z} Zhu, Z.~P., Fu, S.~Y., Liu, X., et al.\ 2021, GRB Coordinates Network, Circular Service, No. 29523, 29523
%\bibitem[MacLachlan, G. A. et al.(2013)]{MacLachlan2013} MacLachlan, G. A., Shenoy, A., Sonbas, E. et al. \ 2013, \mnras, 432, 857-865. doi: 10.1093/mnras/stt241
\end{thebibliography}
\end{document}